\documentclass[%
 reprint,
 amsmath,amssymb,
 aps,
]{revtex4-2}

\usepackage{graphicx}
\usepackage{dcolumn}
\usepackage{bm}
\usepackage[mathlines]{lineno}

\usepackage{amsmath}	
\usepackage{xspace} 
\usepackage{comment}
\usepackage{hyperref}

\newcommand{\be}{\begin{equation}}
\newcommand{\ee}{\end{equation}}

\newcommand{\bi}{\begin{itemize}}
\newcommand{\ei}{\end{itemize}}
\newcommand{\ben}{\begin{enumerate}}
\newcommand{\een}{\end{enumerate}}
\newcommand{\zenodo}{\url{https://zenodo.org/doi/10.5281/zenodo.10458463}}

\newcommand{\onesize}{1.0}

\newcommand{\gcegen}{
\cite{2009arXiv0910.2998G, 2009arXiv0912.3828V, 2011PhLB..697..412H, 2011PhRvD..84l3005H, 2012PhRvD..86h3511A,
2013PhRvD..88h3521G, 2013PDU.....2..118H, 2015JCAP...03..038C, 2016PDU....12....1D, 2016ApJ...819...44A, 2017ApJ...840...43A}\xspace}

\newcommand{\mspmore}{
\cite{2011JCAP...03..010A, 2012PhRvD..86h3511A, 2013PhRvD..88h3521G, 2015JCAP...02..023P, 2015ApJ...812...15B, 2017arXiv170500009F, 2018NatAs...2..387M, 2018NatAs...2..819B, 2018ApJ...863..199G, 2018MNRAS.481.3966B, 2019JCAP...09..042M, 2022JCAP...06..025D}\xspace}

\newcommand{\dmgen}{
\cite{2011PhLB..697..412H, 2011PhRvD..84l3005H, 2012PhRvD..86h3511A, 2013PhRvD..88h3521G, 2013PDU.....2..118H, 2015JCAP...03..038C, 2016PDU....12....1D, 2017ApJ...840...43A}\xspace}

\newcommand{\Fermi}{{\textit{Fermi}}}
\newcommand{\FBs}{{{\Fermi} bubbles}\xspace}
\newcommand{\mnras}{{MNRAS}}
\newcommand{\apjs}{{ApJS}}
\newcommand{\jcap}{{JCAP}}

\begin{document}


\title{Towards resolving the Galactic center GeV excess with\\ millisecond-pulsar-like sources using machine learning}

\author{Dmitry V. Malyshev}
\affiliation{Erlangen Centre for Astroparticle Physics, \\
Nikolaus-Fiebiger-Str. 2, Erlangen 91058, Germany}
 \email{dmitry.malyshev@fau.de}


\date{\today}

\begin{abstract}
Excess of gamma rays around the Galactic center (GC) observed in the {\Fermi} Large Area Telescope (LAT) data is one of the most intriguing features in the gamma-ray sky.
The spherical morphology and the spectral energy distribution with a peak around a few GeV are consistent with emission from annihilation of dark matter particles.
Other possible explanations include a distribution of millisecond pulsars (MSPs).
One of the caveats of the MSP hypothesis is the relatively small number of associated MSPs near the GC.
In this paper, we perform a multiclass classification of \Fermi-LAT sources using machine learning and 
determine the contribution from unassociated MSP-like sources near the GC.
The spectral energy distribution, spatial morphology, and the source count distribution are consistent with expectations for a population of MSPs that can explain the gamma-ray excess.
Possible caveats of the contribution from the unassociated MSP-like sources are discussed.
\end{abstract}

\maketitle



\section{Introduction}

The origin of excess of GeV gamma rays near the Galactic center (GC) \gcegen is one of the most important
questions in high-energy astrophysics.
The excess has an approximately spherical morphology around the GC 
and a maximum in spectral energy distribution (SED) around a few GeV,
which gives a basis for interpretation in terms of dark matter (DM) annihilation \dmgen.
Astrophysical explanations include a population of  millisecond pulsars (MSPs)
\mspmore or
young pulsars \cite{2015arXiv150402477O}.
In general, statistical methods have been used to distinguish a contribution from a population
of point sources below the detection threshold from a smooth DM annihilation profile 
\cite{2015JCAP...05..056L, 2016PhRvL.116e1102B, 2016PhRvL.116e1103L, 2020PhRvL.124w1103Z, 2020PhRvL.125x1102L, 2021PhRvD.104l3022L, 2022PhRvD.105f3017M, 2023JCAP...06..013C}.
Such statistical methods are prone to uncertainties in background diffuse emission,
which can make a signal from DM annihilation look like a distribution of 
pointlike sources
\cite{2019PhRvL.123x1101L, 2020PhRvL.125l1105L, 2020PhRvD.102f3019L}.
As a result, a careful assessment of diffuse background uncertainties is necessary 
\cite{2020PhRvD.101b3014C, 2020PhRvD.102b3023B, 2020PhRvL.125x1102L, 2021PhRvL.127p1102C, 2021PhRvD.104l3022L, 2022PhRvD.105f3017M,
2023JCAP...06..013C}.

The spectrum and morphology of the Galactic center excess (GCE) are also sensitive to additional sources of cosmic ray electrons 
or protons near the GC \cite{2014JCAP...10..052P, 2015JCAP...12..005C, 2016PhRvL.117k1101C, 2017ApJ...840...43A}.
Furthermore, it has been argued that the GCE template proportional to a stellar bulge near the GC
fits the data better than a spherically symmetric DM annihilation profile \cite{2018NatAs...2..387M, 2018NatAs...2..819B, 2019JCAP...09..042M, 2022ApJ...929..136P, 2024PhRvD.109l3042M}.
This statement was, however, challenged in 
Refs.~\cite{2021PhRvD.103f3029D, 2022PhRvD.105j3023C, 2023MNRAS.522L..21M, 2024PhRvD.109l3017Z},
where it was shown that a spherically symmetric DM annihilation profile is preferred over stellar bulge templates.
Overall, the inferred GCE morphology and spectrum as well as interpretation in terms of
faint point sources vs DM annihilation sensitively depend on the uncertainties of the foreground gamma-ray emission.

Although uncertainties in the diffuse emission background prevent an immediate unambiguous solution of the GCE problem
in terms of either a population of faint sources or DM annihilation,
one of the testable predictions of the point sources hypothesis is that deeper observations with the {\Fermi} satellite should lead to
more resolved MSP-like sources near the GC.
This gradual resolution of point sources should lead to a decrease of the remaining excess.
In particular in the MSP hypothesis, the newer \Fermi-LAT catalogs should contain more MSPs (or MSP-like) sources near the GC relative to the older catalogs
 (one can also perform searches for new MSPs in, e.g., radio observations \cite{2016ApJ...827..143C, 2018ApJ...863..199G}).
The lack of the associated MSPs near the GC has been one of the main arguments against the MSP hypothesis.
It has been shown that the number of associated MSPs near the GC is smaller than 
what is expected for a population of MSPs that, on the one hand, can explain the GeV excess and, on the other hand,
is consistent with the populations of MSPs observed locally and in globular clusters \cite{2016JCAP...08..018H}.
Also the number of low-mass X-ray binaries (LMXBs), which are progenitors of MSPs, observed near the GC
is smaller than expectations based on observations of LMXBs in globular clusters \cite{2017JCAP...05..056H}.

Apart from associated \Fermi-LAT sources, there are many unassociated sources near the GC.
In general, there is a gap between studies, which use associated sources
to infer the possible population of MSPs near the GC, and statistical methods, which determine the total population of pointlike sources in the region, 
including unassociated sources and sources below the detection threshold, but which do not distinguish different classes of sources (for a discussion see Ref.~\cite{2024arXiv240603990M}).
The main purpose of this paper is to partially close this gap and to perform a multiclass classification of \Fermi-LAT sources with machine learning algorithms
in order to estimate the number and the distribution of MSP-like sources among the unassociated gamma-ray sources near the GC.

The paper is organized as follows.
In Sec.~\ref{sec:multiclass} we perform a multiclass classification of \Fermi-LAT sources using machine learning methods.
In Sec.~\ref{sec:msp_distr} we determine the expected contribution of unassociated MSP-like sources to the SED within $10^\circ$ from the GC, 
to the radial intensity profile at 2 GeV, and to the source count distribution within $10^\circ$ from the GC.
Sec.~\ref{sec:conclusion} contains conclusions and discussion.
In Appendix~\ref{app:msp_syst} we discuss the systematic uncertainties of the expected contribution of MSP-like sources,
while the contribution of young pulsars among the unassociated sources is discussed in Appendix~\ref{app:psr_contribution}.

\section{multiclass classification of \Fermi-LAT sources}
\label{sec:multiclass}

In order to determine the contribution of MSP-like sources among the unassociated \Fermi-LAT sources to the gamma-ray flux near the GC,
we perform multiclass classification of sources in the fourth \Fermi-LAT catalog data release four 
(4FGL-DR4 \cite{2022ApJS..260...53A, 2023arXiv230712546B} version ``gll\_psc\_v34.fit'').
The classification follows the procedure developed in Refs.~\cite{2023MNRAS.521.6195M, 2023RASTI...2..735M}
with the following two modifications:
\ben
\item
In order to avoid a possible bias of the classification near the GC,
we exclude coordinate features (Galactic latitude and longitude) from the list of input features.
As a result, we use the following seven input features
based on the source parameters in the 4FGL-DR4 catalog \citep{2022ApJS..260...53A, 2023arXiv230712546B}:
$\log_{10}$(Energy\_Flux100), 
$\log_{10}$(Unc\_Energy\_Flux100), 
$\log_{10}$(Signif\_Avg), 
LP\_beta,
LP\_SigCurv,
$\log_{10}$(Variability\_Index),
and the index of the log parabola spectrum at 1 GeV.
For a discussion about the selection of features, please see Refs.~\cite{2020MNRAS.492.5377L, 2022A&A...660A..87B, 2023MNRAS.521.6195M, 2023RASTI...2..735M}.
We also show in Appendix~\ref{app:msp_syst} that inclusion of source coordinates as input features has little effect on the predictions for 
MSP-like sources near the GC.
\item
We exclude from training associated sources with uncertain associations, 
such as blazars of unknown type (bcu) and sources with both a supernova remnant and a pulsar wind nebula in the vicinity (spp) in addition to sources with unknown nature of the association counterpart (unk). 
The classes used for the training are summarized in the first section of Table~\ref{tab:classes}.
The associated classes, which are not used for training in the main part of the paper are in the second section of that table.
The effect of inclusion of bcu and spp sources in training is discussed in Appendix~\ref{app:msp_syst}.
\een

\begin{figure}
\centering
\includegraphics[width=\onesize\columnwidth]{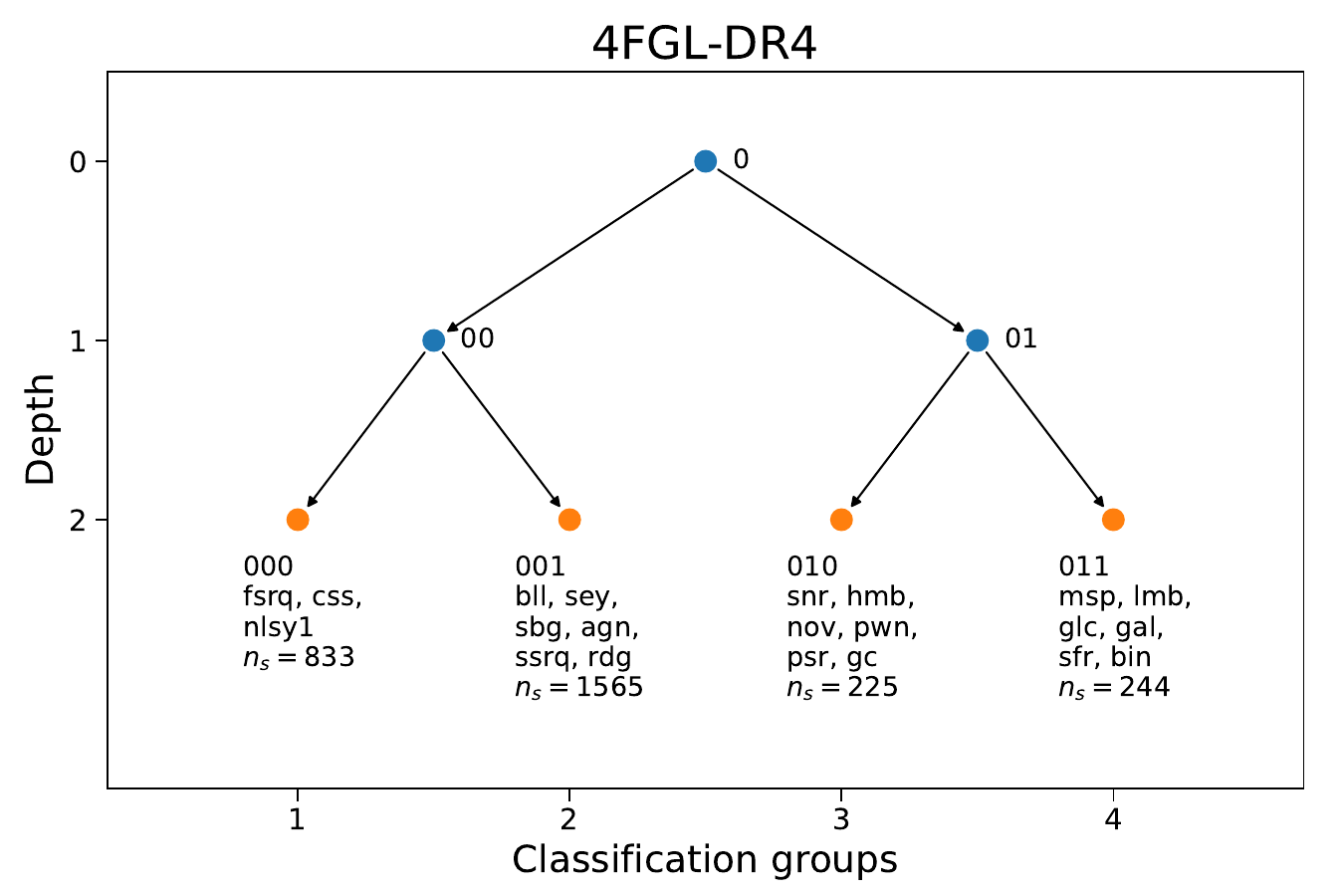}
\caption{
Hierarchical definition of classes following the procedure in Refs.~\cite{2023MNRAS.521.6195M, 2023RASTI...2..735M}.
The classes used for the classification in this work are shown at the bottom row (depth 2). The numbers at the bottom show the total count of associated sources in the group.}
\label{fig:class_def}
\end{figure}

For the definition of the classes, we separate the physical classes into groups using the hierarchical procedure 
\cite{2023MNRAS.521.6195M} based on the Gaussian mixture model.
We require that there are not fewer than 100 sources in a class.
After two splits, the procedure converges to four classes shown in Fig.~\ref{fig:class_def} at the bottom row (depth 2).

\begin{table}
\centering
\footnotesize
\begin{tabular}{lll}
\hline
Physical class & Count & Description \\
\hline
gc & 1 &  Galactic center\\
psr & 141 &  young pulsar\\
msp & 179 &  millisecond pulsar\\
pwn & 21 &  pulsar wind nebula\\
snr & 45 &  supernova remnant\\
glc & 34 &  globular cluster\\
sfr & 6 &  star-forming region\\
hmb & 11 &  high-mass binary\\
lmb & 9 &  low-mass binary\\
bin & 10 &  binary\\
nov & 6 &  nova\\
bll & 1490 &  BL Lac type of blazar\\
fsrq & 819 &  FSRQ type of blazar\\
rdg & 53 &  radio galaxy\\
agn & 8 &  non-blazar active galaxy\\
ssrq & 2 &  steep spectrum radio quasar\\
css & 6 &  compact steep spectrum radio source\\
nlsy1 & 8 &  narrow-line Seyfert 1 galaxy\\
sey & 4 &  Seyfert galaxy\\
sbg & 8 &  starburst galaxy\\
gal & 6 &  normal galaxy (or part) \\ 
\hline
unk & 147 &  sources with an unknown nature  \\
&& of the association counterpart\\
bcu & 1625 &  blazar candidate of uncertain type\\
spp & 124 &   supernova remnant and/or \\
 & &  pulsar wind nebula \\
 \hline
 unas & 2428 & unassociated \\
\hline
\end{tabular}
\caption{Physical classes of sources in the 4FGL-DR4 catalog \citep{2022ApJS..260...53A, 2023arXiv230712546B}
version ``gll\_psc\_v34.fit''.
Both associated and identified sources in the catalog are referred to as associated sources in this work.
The first part shows the classes used for training. The second part shows the associated sources with uncertain classification.
These sources are not used for training.
The last part gives the number of unassociated sources.
}
\label{tab:classes}
\end{table}

\begin{table*}
\centering
\footnotesize
\begin{tabular}{llllllll}
\hline
 & Class label & Physical classes & N assoc & RF assoc & NN assoc & RF unas+ & NN unas+ \\
\hline
1 & fsrq+ & fsrq, nlsy1, css & 833 & $835.2\pm 24.4$ & $832.9\pm 40.6$ & $1156.1\pm 28.3$ & $1276.2\pm 84.2$ \\
2 & bll+ & bll, sey, sbg, agn, ssrq, rdg & 1565 & $1568.0\pm 31.1$ & $1560.5\pm 43.2$ & $1791.7\pm 35.1$ & $1801.1\pm 66.4$ \\
3 & psr+ & snr, hmb, nov, pwn, psr, gc & 225 & $214.5\pm 18.8$ & $227.8\pm 36.6$ & $576.8\pm 43.4$ & $494.1\pm 97.2$ \\
4 & msp+ & msp, lmb, glc, gal, sfr, bin & 244 & $249.0\pm 17.7$ & $245.5\pm 26.2$ & $799.3\pm 39.1$ & $752.5\pm 64.4$ \\
\hline
\end{tabular}
\caption{Definition of classes and prediction for the expected number of class members among associated and unassociated sources. The ``unas+'' columns include both unassociated sources and sources with uncertain classification (bcu, spp, and unk classes).
The table includes predictions based on RF and NN classification algorithms.}
\label{tab:pred_no_bcu_spp}
\end{table*}

\begin{table*}
\centering
\footnotesize
\begin{tabular}{lllllllll}
\hline
Class label & RF unas & NN unas & RF unk & NN unk & RF bcu & NN bcu & RF spp & NN spp \\
\hline
fsrq+ & $573.7\pm 21.1$ & $652.6\pm 56.6$ & $16.3\pm 1.2$ & $23.8\pm 3.3$ & $552.6\pm 8.2$ & $577.1\pm 24.2$ & $13.5\pm 1.3$ & $22.7\pm 3.5$ \\
bll+ & $834.0\pm 24.8$ & $846.2\pm 39.4$ & $47.6\pm 1.9$ & $55.4\pm 3.0$ & $891.3\pm 9.8$ & $871.6\pm 24.4$ & $18.7\pm 1.6$ & $27.9\pm 2.7$ \\
psr+ & $408.7\pm 33.1$ & $347.6\pm 69.9$ & $47.4\pm 3.0$ & $36.9\pm 6.3$ & $62.1\pm 5.9$ & $67.7\pm 15.8$ & $58.6\pm 3.2$ & $41.9\pm 7.7$ \\
msp+ & $611.5\pm 30.5$ & $581.5\pm 47.1$ & $35.6\pm 2.6$ & $30.9\pm 3.3$ & $119.0\pm 5.6$ & $108.6\pm 13.5$ & $33.2\pm 3.1$ & $31.5\pm 3.7$ \\
\hline
\end{tabular}
\caption{Predictions for the expected number of class members calculated separately for unassociated, unk, bcu, and spp sources.}
\label{tab:pred_unas_bcu_spp_unk}
\end{table*}

In this paper we use two classification algorithms: 
\ben
\item
Random forest (RF) implemented in scikit-learn \citep{scikit-learn} (version 1.2.2).
The maximal number of trees is 50 with the maximal depth 15. 
For the other parameters we use the default values in the scikit-learn implementation of RF.
The RF method is used as a baseline model in the main part of the paper.
\item
Neural networks (NN)  implemented in TensorFlow \citep{tensorflow2015-whitepaper}.
We use stochastic gradient descent (adam) with learning rate of 0.001, two hidden layers with 20 and 10 nodes respectively, 
tanh activation functions, 2000 epochs, batch size of 200, L2 regularization with $l2 = 0.001$, and no drop out.
We use the sparse categorical cross entropy loss function
\footnote{\url{https://www.tensorflow.org/api_docs/python/tf/keras/losses/SparseCategoricalCrossentropy}}.
Comparison of NN and RF predictions are shown in Appendix~\ref{app:msp_syst}.
\een
We use 70/30\% splits into training and testing data samples
and derive the class probabilities for all sources (both associated and unassociated).
The RF and NN metaparameters are selected based on the previous studies 
\cite{2022A&A...660A..87B, 2023RASTI...2..735M}
to ensure that the algorithms are sufficiently flexible, i.e., further increase in the number of free parameters does not lead 
to improvement of performance, and do not overfit the data.
The class probabilities of associated sources are determined by averaging over
training/testing splits, when the sources are in testing samples.
We require that each associated source appears in the testing samples at least 5 times,
which results in 47 training/testing splits.
The class probabilities for the unassociated sources (as well as for the unk, bcu, and spp sources, which are not included in the training) are determined by the average over the 47 realizations of the training/testing splits.
In the probabilistic catalog we report the class probabilities determined with RF and NN algorithms as well as the 
corresponding statistical uncertainties derived as a sample average over 47 realizations of training/testing splits for unassociated sources or, for associated sources, as an average over the splits when the source is included in the testing sample.
The corresponding catalog is available online at \zenodo.

One of the important questions is the reliability of the classification in the presence of possible misreconstruction of source parameters, e.g., due to mismodeling of diffuse background,
as well as the differences in the distribution of associated and unassociated sources.
In this paper, we will assume that the differences in the distribution of associated and unassociated sources is due to covariate 
shift~\cite{2023RASTI...2..735M}.
Both mismodeling of sources and the differences in the distribution of associated and unassociated sources
can be tested, e.g., by comparing the classification of unassociated sources in an earlier catalog, e.g., 3FGL \cite{2015ApJS..218...23A},
with the new associations in a later catalog, e.g., 4FGL \cite{2020ApJS..247...33A, 2020arXiv200511208B}, which also utilizes an updated diffuse emission model.
The drop in accuracy is found to be between 3\% and 10\% depending on the classification algorithms and the number of classes
\cite{2020MNRAS.492.5377L, 2021RAA....21...15Z, 2021MNRAS.507.4061F, 2022A&A...660A..87B}.
Alternatively, if no newer catalogs are available, one can estimate the effect of covariate shift
by reweighting the associate sources in the testing sample so that the weighted distribution of associated sources is similar to the distribution of unassociated sources
\cite{2023RASTI...2..735M}.
The effect of the weighted training and testing can be seen in the confusion matrices in Figs.~14 and C4 of Ref.~\cite{2023RASTI...2..735M} in the case of six-class classification.
The numbers on the diagonals show the precision of predictions (i.e., the fraction of true sources in the class relative to the number of predicted class members).
For some classes, the drop in precision can be as high as 20\% for weighted vs unweighted testing, e.g., for BL Lacs. For the msp+ class, the corresponding drop is not very significant.
In this work, we compare the expected spectrum from MSP-like sources for weighted vs unweighted training and testing in Appendix \ref{app:msp_syst}, Fig.~\ref{fig:msp_many_cats}.
The corresponding effect is less than about 20\% for the RF algorithm.
For NN, the effect is also less than 20\% around a few GeV but increases up to about 50\% around 100 MeV and 100 GeV.

The number of unassociated sources in class $m$ can be estimated by summing the corresponding class probabilities
\be
\label{eq:ntot}
N_m = \sum_{i \in {\rm unas}} p_{m}^i.
\ee
Similarly, one can estimate the number of associated sources in class $m$ (and compare with the actual counts of associated sources in this class) or the number of class-$m$ sources among unk, bcu, or spp sources.
In Table~\ref{tab:pred_no_bcu_spp} we show the expected numbers of class members.
In this table we add the predictions for the unas, unk, bcu, and spp sources together.
The predictions for the unas, unk, bcu, and spp sources separately are presented in Table~\ref{tab:pred_unas_bcu_spp_unk}.
The statistical uncertainty on $N_m$ is determined from the sample variance for the probabilities determined with the different test sample choices.
The expected number of sources in a class among associated sources is computed analogously to the unassociated sources, 
where the sum in Eq.~(\ref{eq:ntot}) goes over all associated rather than unassociated sources.
As one can see in Table~\ref{tab:pred_no_bcu_spp}, the predicted numbers of class members among associated sources are consistent with the actual source counts
of associated sources.
RF and NN algorithms agree on the predicted numbers of bll+ and fsrq+ sources among the unassociated sources.
Although the predictions for psr+ and msp+ classes have relatively large differences, the predictions are still within the statistical errors.

\section{Distribution of unassociated MSP-like sources}
\label{sec:msp_distr}

\subsection{Energy spectrum}

\begin{figure}
\centering
\includegraphics[width=\onesize\columnwidth]{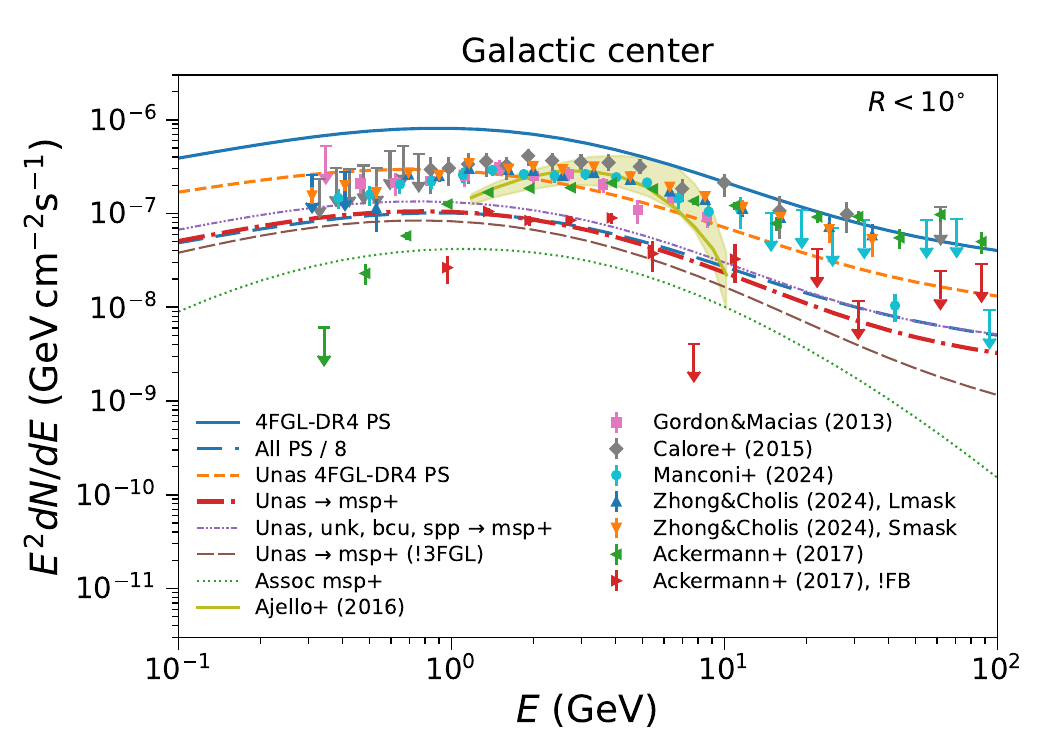}
\caption{
Comparison of the GCE SED with the SED from MSP-like sources within $10^\circ$ from the GC.
Blue solid (long-dashed) line -- combined SED (combined SED divided by 8) for all 4FGL-DR4 sources within $10^\circ$ from the GC,
orange dashed line -- combined SED of all unassociated sources in the ROI,
red dash-dotted line -- the expected contribution from msp+ sources among the unassociated sources,
purple dash-dot-dotted line -- the expected contribution from msp+ sources among the unassociated, unk, bcu, spp sources,
brown dashed line -- the expected contribution from msp+ sources among the unassociated sources after excluding the sources detected in the 3FGL catalog,
green dotted line -- SED for the associated MSPs.
The GCE models include 
Gordon \& Macias~(2013) \cite{2013PhRvD..88h3521G},
Calore+ (2015)~\cite{2015JCAP...03..038C},
Ajello+ (2016)~\cite{2016ApJ...819...44A},
Ackermann+ (2017)~\cite{2017ApJ...840...43A},
Manconi+ (2024)~\cite{2024PhRvD.109l3042M}, and
Zhong \& Cholis (2024)~\cite{2024PhRvD.109l3017Z}.
See text for more details about the GCE models.
}
\label{fig:msp_gc_sp}
\end{figure}

In this section we derive the expected combined SED
that can be attributed to the msp+ class near the GC.
We consider sources within $10^\circ$ from the GC and determine the SED
by summing the log-parabola spectra weighted by the msp+ class probabilities.
For example, the expected contribution of the msp+ class to the SED of unassociated sources
is determined as
\be
\label{eq:SED}
F_{\rm msp+}(E) = \sum_{i\in {\rm unas}} p_{\rm msp+}^i F_i(E),
\ee
where $p_{\rm msp+}^i $ is the msp+ class probability for source $i$ and $F_i(E)$ is the SED for source $i$ modeled by the log-parabola spectrum.
The corresponding msp+ SED is shown in Fig.~\ref{fig:msp_gc_sp} by the red dash-dotted line (labeled as ``Unas $\rightarrow$ msp+'').
If we sum over all unas, unk, bcu, and spp sources in Eq.~(\ref{eq:SED}), then the expected SED is shown by the purple dash-dot-dotted line (labeled as ``Unas, unk, bcu, spp $\rightarrow$ msp+'').
We see that the effect of
including unk, bcu, spp sources in addition to the unassociated ones is relatively small.

The contribution of the 4FGL-DR4 unassociated sources, which do not have a counterpart in the 3FGL catalog,
is shown by the brown dashed line (labeled as ``Unas $\rightarrow$ msp+ (!3FGL)'').
This is the emission, which was previously not resolved in the 3FGL catalog and must have been 
attributed to a diffuselike component, e.g., the GCE.
For comparison, the combined SED of all resolved sources in the region of interest (ROI) is shown by the blue solid line.
We also show by the blue long-dashed line the combined SED of all resolved sources divided by 8 in order to match the SED of msp+ sources around 1 GeV.
The SED of all sources is harder above about 10 GeV due to the contribution of blazars.
The SED of all unassociated sources in the ROI is shown by the orange dashed line.
The SED of all associated msp+ sources is shown by the green dotted line.
We compare the msp+ SED with the GCE SEDs from Refs.~\citep{2013PhRvD..88h3521G, 2015JCAP...03..038C, 2016ApJ...819...44A, 2017ApJ...840...43A, 2024PhRvD.109l3042M, 2024PhRvD.109l3017Z}. We note that different analyses of GCE have slightly different ROIs. Provided that we are interested in a qualitative agreement, we take the ROIs which are closest to the circle of $10^\circ$ radius around the GC and apply no correction for the differences in the ROI.
For the analysis of Ajello {\it et al.} (2016)~\cite{2016ApJ...819...44A} (yellow band), 
we use the model with CR sources traced by pulsars, where both the normalization and the index of the background components are fitted.
For Zhong \& Cholis (2024)~\cite{2024PhRvD.109l3017Z}, we use Model I with large and small masks around point sources labeled by Lmask and Smask respectively.
The labels ``Ackermann+ (2017), !FB'' and ``Ackermann+ (2017)'' correspond to models in Ref.~\cite{2017ApJ...840...43A} that
that take and do not take into account the contribution from \FBs near the GC respectively.
It is interesting, that the contribution of the msp+ class to the unassociated sources (also when sources with the 3FGL counterparts are not included) 
is comparable around a few GeV to the SED attributed to the GCE in a model that removes a possible emission from the  \FBs at low latitudes.

\begin{figure}
\centering
\includegraphics[width=\onesize\columnwidth]{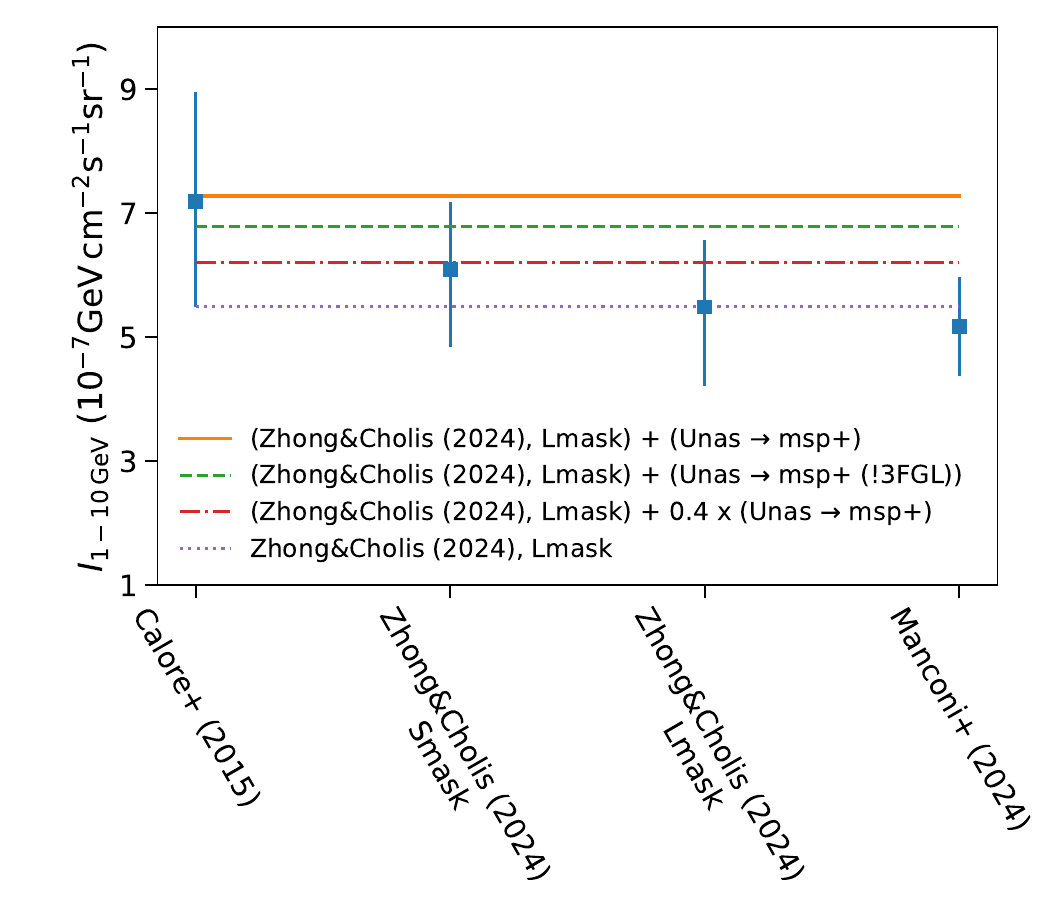}
\caption{
Energy flux of GCE integrated between 1 GeV and 10 GeV for the GCE models from 
Calore {\it et al.} (2015)~\cite{2015JCAP...03..038C},
Zhong\&Cholis (2024)~\cite{2024PhRvD.109l3017Z} Model I with small (Smask) and large (Lmask) masks around point sources, 
and Manconi {\it et al.} (2024)~\cite{2024PhRvD.109l3042M}.
Dotted purple line corresponds to large-mask model of Ref.~\cite{2024PhRvD.109l3017Z}.
Red dash-dotted line represents adding 40\% of msp+ contribution to unassociated sources
to the large-mask model of Ref.~\cite{2024PhRvD.109l3017Z}.
Solid orange (dashed green) line represents adding the msp+ contribution to unassociated sources
to the large-mask model of Ref.~\cite{2024PhRvD.109l3017Z} with (without) 3FGL sources.
See text for more details.
}
\label{fig:W_evolv}
\end{figure}

In Fig.~\ref{fig:W_evolv}, we compare the energy spectrum of msp+ sources integrated between 1 GeV and 10 GeV with the GCE in 
Refs.~\cite{2015JCAP...03..038C, 2024PhRvD.109l3042M, 2024PhRvD.109l3017Z}.
In this plot, we use the same ROI as in Refs.~\cite{2015JCAP...03..038C, 2024PhRvD.109l3042M}: $|\ell| < 20^\circ$ and $2^\circ < |b| < 20^\circ$.
The x axis shows the different models: Calore {\it et al.} (2015)~\cite{2015JCAP...03..038C}, Zhang \& Cholis (2024)~\cite{2024PhRvD.109l3017Z} with small (Smask) and large (Lmask) masks around point sources, and Manconi {\it et al.} (2015)~\cite{2024PhRvD.109l3042M}.
We note that the model in Ref.~\cite{2024PhRvD.109l3042M} does not have a mask around the Galactic plane, i.e., the ROI is 
$|\ell| < 20^\circ$ and $|b| < 20^\circ$.
The points are obtained by integrating the interpolated SEDs in Refs.~\cite{2015JCAP...03..038C, 2024PhRvD.109l3042M, 2024PhRvD.109l3017Z}.
The error bars represent the integrated upper and lower bands of the corresponding models.
In order to compare the different models we show the model in Ref.~\cite{2024PhRvD.109l3017Z} with the large mask as the dashed purple line.
The small (large) mask of Ref.~\cite{2024PhRvD.109l3017Z} corresponds to approximately 50\% (90\%)
containment~\footnote{\url{https://www.slac.stanford.edu/exp/glast/groups/canda/lat_Performance.htm}}
for the front-converting events used in the analysis of Ref.~\cite{2024PhRvD.109l3017Z}.
Consequently, the expected additional contribution of msp+ sources in the small-mask case relative to the large-mask case is about 40\% of the msp+ sources.
The sum of the large-mask model and the 40\% of the msp+ contribution to the unassociated sources is shown by the red dash-dotted line (here we neglect the contribution of the associated msp+ sources).
It is consistent with the small-mask model of Ref.~\cite{2024PhRvD.109l3017Z}.
The sums of the large-mask model and the msp+ contribution to the unassociated sources including or excluding 3FGL sources are shown by the solid orange and dashed green lines respectively.
We see that the expected contribution of msp+ sources is also consistent with an earlier estimate of the GCE in Ref.~\cite{2015JCAP...03..038C}, while a later model in Ref.~\cite{2024PhRvD.109l3042M} that includes a contribution of point sources near the GC corresponds to a further reduction of the flux attributed to GCE.
We note that the templates and background models in Refs.~\cite{2015JCAP...03..038C, 2024PhRvD.109l3042M, 2024PhRvD.109l3017Z}
are very different, i.e., a direct comparison is complicated due to large systematic uncertainties (included in the error bars around the models).
The best way to see a direct contribution of msp+ sources is to compare predictions of the same model (e.g., Model I in Ref.~\cite{2024PhRvD.109l3017Z}) with different masks around point sources.
Nevertheless, we find that, in general, the reduction of the flux attributed to the GCE is consistent both with the inclusion of the contribution of point sources in the later \Fermi-LAT catalogs and with the contribution of MSP-like sources among unassociated \Fermi-LAT sources.

\subsection{Radial profile}

\begin{figure}
\centering
\includegraphics[width=\onesize\columnwidth]{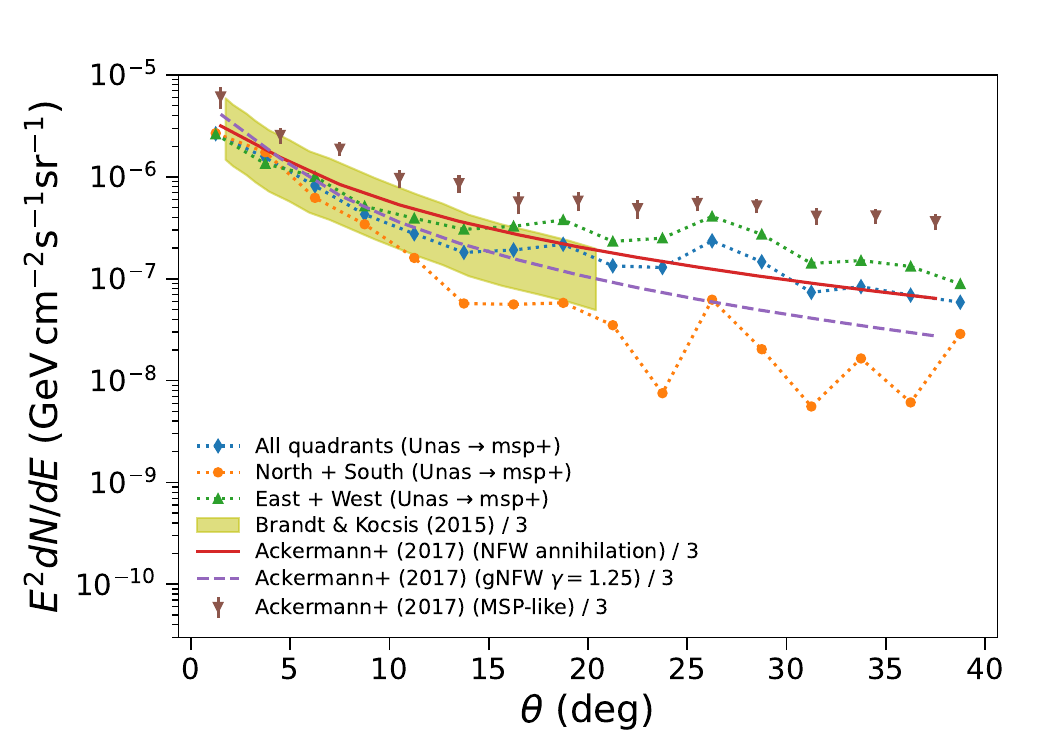}
\caption{
Radial profile of expected intensity from msp+ sources at 2 GeV defined in Eq.~(\ref{eq:SED}).
Blue diamonds, orange circles, and green upward triangles show respectively the intensity in all, north+south, east+west quadrants.
For comparison, we show the GCE profiles (divided by 3)
from Brandt \& Kocsis (2015) \cite{2015ApJ...812...15B} and Ackermann {\it et al.} (2017) \cite{2017ApJ...840...43A}. 
See text for more details.
}
\label{fig:radial}
\end{figure}

We compare in Fig.~\ref{fig:radial} the radial profile of the unassociated msp+ component to the radial profiles of GCE models at 2 GeV.
The average intensity of the msp+ component of the unassociated sources in a ring between $\theta_1$ and  $\theta_2$ from the GC 
is estimated as 
\be
\label{eq:SED}
I_{\rm msp+}(\theta) = \frac{1}{\Omega} \sum_{i\in {\rm unas},\ \theta_i \in (\theta_1,\ \theta_2)} p_{\rm msp+}^i F_i(E = 2\ {\rm GeV}),
\ee
where $\theta = (\theta_1 + \theta_2) / 2$ is the mean radius and $\Omega$ is the solid angle of the ring.
The radial profile of the expected intensity of emission at 2 GeV for the msp+ component among the unassociated sources
is shown by the blue diamonds in Fig.~\ref{fig:radial}.
We also separately plot the intensities in the north+south quadrants (orange circles) and east+west quadrants (green upward triangles).
The quadrants are determined by comparing the $z$ and $y$ projections of the unit vectors on the sphere, where $x$ goes from the observer to the GC, while $y$ and $z$ are, respectively, in the positive Galactic longitude and latitude directions from the GC.
The quadrants are obtained by splitting the sphere by two planes: $z = y$ and $z = -y$.
For instance, the northern quadrant is determined by the condition $z > abs(y)$.

In order to facilitate the comparison of the radial profile of MSP-like sources with models of the GCE,
we divide the models listed below by 3.
The model for the MSP population derived from disrupted globular clusters \cite{2015ApJ...812...15B}
is shown by the yellow band.
We also consider several models from Ref.~\cite{2017ApJ...840...43A}: the NFW DM annihilation profile (solid red line), the generalized NFW DM annihilation profile with index $\gamma = 1.25$ (dashed purple line), and a model of the GCE derived from a correlation with MSP-like energy spectrum (Fig.~11 of Ref.~\cite{2017ApJ...840...43A}) shown by the brown downward triangles.

\begin{figure*}
\centering
\includegraphics[width=1.7\columnwidth]{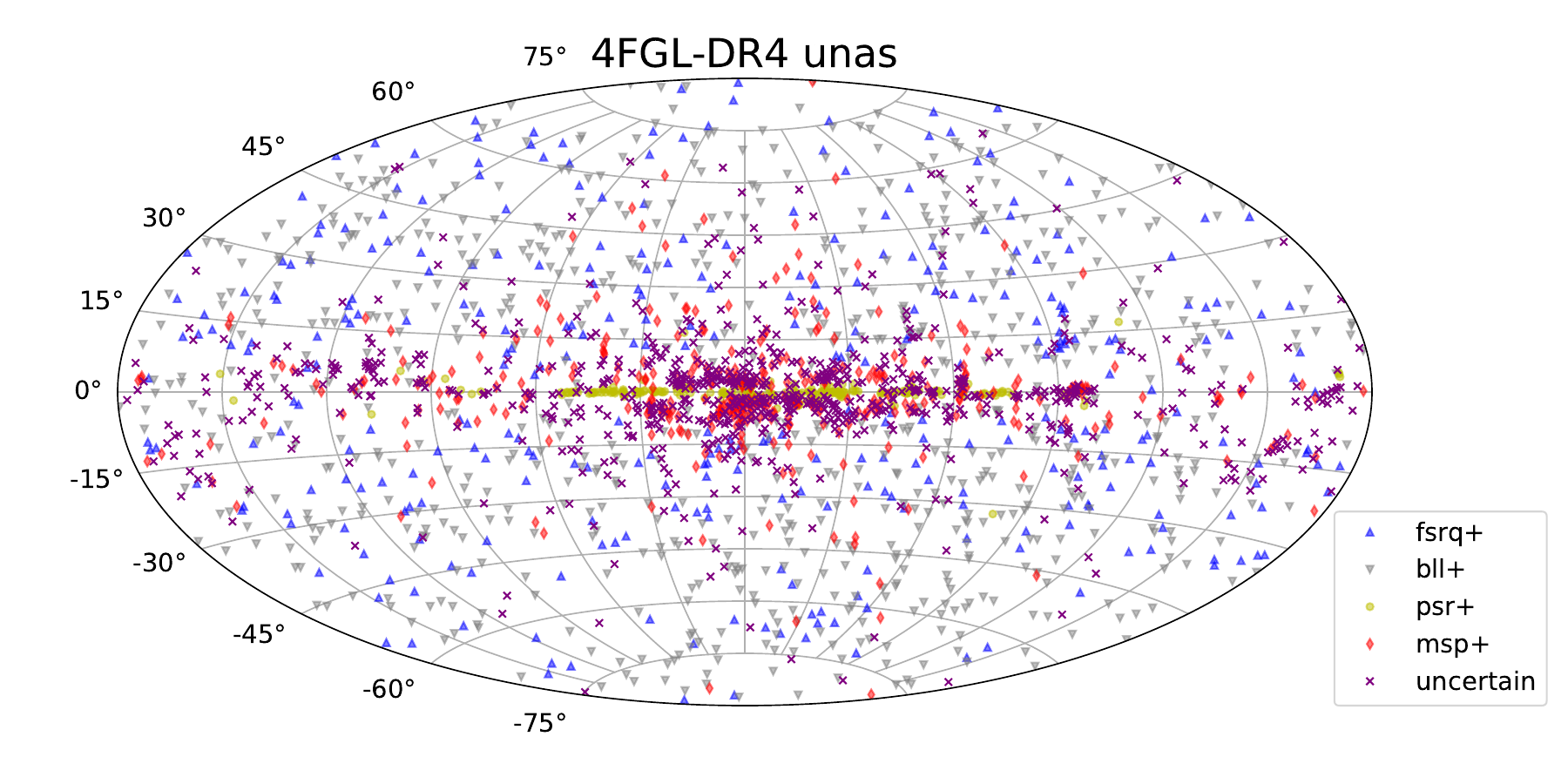}
\caption{
Distribution of class candidates among unassociated sources. 
An unassociated source is attributed to a class if the corresponding class probability is larger than 0.5.
If all class probabilities are smaller than 0.5, then the sources is labeled as ``uncertain''.
}
\label{fig:map}
\end{figure*}

The average profile and the north+south, east+west profiles have a reasonable agreement with the NFW (or gNFW) DM annihilation profiles within about $10^\circ$ from the GC.
At larger separations, the east+west (north+south) distribution is flatter (steeper) than the NFW profile.
We interpret this difference as a contribution of two different populations of MSPs: the bulge population that has an approximately spherical distribution around the GC, which dominates the intensity within $10^\circ$ from the GC, and a disk component, which dominates the emission for the east+west quadrants further away from the GC and is absent in the north+south quadrants at large angles from the GC.
Provided that the radial profiles in east+west and north+south quadrants start to deviate above $10^\circ$ from the GC, we estimate that the width of the disc component is about $10^\circ$ above and below the Galactic plane 
(this is also consistent with a relatively wide distribution of MSP-like sources around the Galactic plane in Fig.~\ref{fig:map}). 
The difference in intensity between east+west and north+south quadrants around $10^\circ - 20^\circ$ from the GC can serve as an estimate of the intensity of the disc component of MSPs $E^2 dN_{\rm disc}/dE \sim 3\times 10^{-7}\ {\rm GeV\ cm^{-2} s^{-1} sr^{-1}}$.
Thus, the disc component of MSPs can provide about 15\% contribution to intensity from MSP-like sources near the GC or
about 30\% contribution around $5^\circ$ from the GC.

It is interesting to note that the intensity of excess emission derived from the correlation with the MSP-like energy spectrum 
(the brown downward triangles in Fig.~\ref{fig:radial})
is generally higher than the other GCE profiles (cf. Ref.~\cite{2017ApJ...840...43A}), it is also higher than the distribution of the msp+ component among the unassociated sources (even after dividing by the factor of 3).
This may be due to a further contribution from an unresolved population of MSPs.
In addition, there is potentially an important contribution from young pulsars (Appendix~\ref{app:psr_contribution}).

The distribution of class candidates over the whole sky is shown in Fig.~\ref{fig:map}.
Although coordinates of the sources were not used as input features, the general distributions of class candidates are reasonable.
In particular, fsrq+ and bll+ candidates have uniform distributions on the sky,
the psr+ candidates (which also include SNRs and PWNe) are distributed along the Galactic plane, while the msp+ candidates (which also include globular clusters) have a wide distribution around the Galactic plane.

\subsection{Source count distribution}

\begin{figure}
\centering
\includegraphics[width=\onesize\columnwidth]{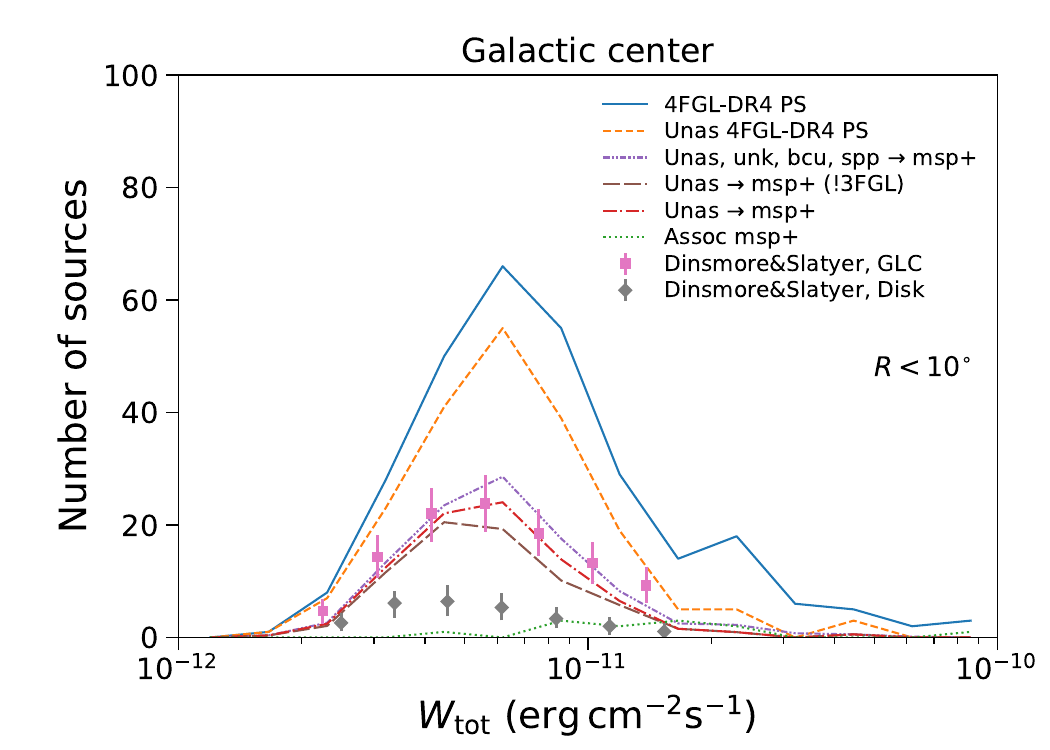}
\caption{
Distribution of the expected source count for the msp+ component of unassociated sources as a function of 
energy flux integrated above 100 MeV.
Blue solid (orange dashed) line -- all (unassociated) sources in the 4FGL-DR4 catalog detected within $10^\circ$ from the GC,
red dash-dotted (purple dash-dot-dotted) line -- expected msp+ contribution among unassociated (unassociated, unk, bcu, and spp) sources,
brown dashed line -- msp+ contribution to unassociated sources after excluding sources detected in the 3FGL catalog.
Green dotted line -- distribution of associated msp+ sources.
We compare the distributions with two models of MSPs with luminosity functions scaled to reproduce
the GCE (Fig.~9a of Ref.~\cite{2022JCAP...06..025D}):
a model based on the properties of globular clusters \citep{2016JCAP...08..018H} -- pink squares labeled as ``GLC'',
and a model based on the properties of MSPs in the Galactic disk \citep{2018MNRAS.481.3966B} -- gray diamonds labeled as ``Disk''.
}
\label{fig:source_count}
\end{figure}

The number of msp+ sources among the unassociated sources with
integrated energy fluxes $W_i \in (W_1,\ W_2)$ is estimated as
\be
\label{eq:dNdW}
N_{\rm msp+}(W) = \sum_{i\in {\rm unas},\ W_i \in (W_1,\ W_2)} p_{\rm msp+}^i,
\ee
where $W_i$ is the energy flux above 100 MeV for an unassociated source $i$ (we use the ``Energy\_Flux100'' values in the catalog) and $W$ is the logarithmic center of the bin: $W = \sqrt{W_1 W_2}$.
In Fig.~\ref{fig:source_count} we show the source count distribution (for sources within $10^\circ$ from the GC) 
for the msp+ component of unassociated sources (red dash-dotted line), of unassociated, unk, bcu, and spp sources (purple dash-dot-dotted line), of unassociated sources, where the 3FGL sources are excluded (brown dashed line).
We also show the distributions 
(within the $10^\circ$ radius from the GC)
for all 4FGL-DR4 sources (solid blue line), for all unassociated 4FGL-DR4 sources (dashed orange line), and for associated msp+ sources (dotted green line).
We compare the distributions with two models of MSPs with luminosity functions scaled to reproduce
the GCE (Fig.~9a of Ref.~\cite{2022JCAP...06..025D}):
a model based on the properties of globular clusters \cite{2016JCAP...08..018H} -- pink squares labeled as ``GLC'',
and a model based on the properties of MSPs in the Galactic disk \cite{2018MNRAS.481.3966B} -- gray diamonds labeled as ``Disk''.
These two models provide reasonable benchmark cases for many other models of the MSP luminosity functions,
e.g., Refs.~\cite{2016PhRvL.116e1102B, 2016PhRvL.116e1103L, 2018ApJ...863..199G, 2020PhRvL.124w1103Z, 2020JCAP...12..035P, 2022NatAs...6..703G}.
Interestingly, the expected number of msp+ sources among the unassociated sources in the 4FGL-DR4 catalog seem to favor the luminosity function for MSPs derived from the globular clusters \cite{2016JCAP...08..018H}.

We note that the luminosity function of the GLC model \cite{2016JCAP...08..018H} is normalized in Ref.~\cite{2022JCAP...06..025D} to reproduce the full GCE. Although the flux from the MSP-like candidates is about a factor of 3 smaller than the full GCE SED, the expected source count of MSP-like sources  in Fig.~\ref{fig:source_count} is consistent with the expected number of MSPs that can be detected by the {\Fermi} LAT.
The reason for this apparent discrepancy is that, although the luminosity function of the GLC model is dominated by bright sources relative to most of other models of the MSP population near the GC, the peak of the luminosity function in the GLC model is still below the {\Fermi}-LAT flux sensitivity.
In particular, the {\Fermi}-LAT sensitivity is estimated to be about $4\times 10^{-12}\ {\rm erg\,cm^{-2} s^{-1}}$ (Fig.~5 of Ref.~\cite{2022JCAP...06..025D}),
while the peak in $L dN/dL$ distribution is at about $4\times 10^{-13}\ {\rm erg\,cm^{-2} s^{-1}}$ (Fig.~7 of Ref.~\cite{2022JCAP...06..025D}).
Thus, also in the GLC model the majority of the GCE SED is expected to come from MSPs undetected by the {\Fermi} LAT.

\section{Conclusions and discussion}
\label{sec:conclusion}

In this paper, we performed a multiclass classification of \Fermi-LAT sources into four classes dominated by 
fsrq, bll, psr, and msp classes.
We excluded unk, bcu, and spp classes from training of machine learning classification algorithms since the nature of the association counterparts is either unknown or uncertain for these classes.
One of the biggest effects from excluding unk, bcu, and spp classes from training is that these classes are also excluded from the classification of the unassociated sources.
As a result, the expected number of, e.g., MSP-like sources among the unassociated ones is much larger in this case, compared to a classification where bcu and spp sources are included in training and classification
\citep{2023MNRAS.521.6195M, 2023RASTI...2..735M} (Appendix \ref{app:msp_syst}).

We find that a significant fraction, i.e., around 30\% at a few GeV, of GCE is now resolved into point sources consistent with MSP-like sources.
We show in Fig.~\ref{fig:W_evolv} that the explanation of the GCE in terms of MSP-like sources is also consistent with 
analyses, which use masks of different sizes around detected sources~\cite{2020PhRvL.124w1103Z, 2022PhRvD.105j3023C, 2024PhRvD.109l3017Z}.
Although this result shows that the problem of the GCE can be resolved by detecting individual gamma-ray sources near the GC,
there are several caveats in the analysis, which require a further investigation:
\ben
\item
The spectrum of the msp+ component is relatively flat below 1 GeV while the spectrum of GCE determined in some analyses, including the analysis that takes the FB into account \citep{2017ApJ...840...43A},
is hard below 1 GeV. The GCE spectrum at low energies has large systematic uncertainties, i.e., it can also be relatively soft \cite{2013PhRvD..88h3521G}. 
This question can be resolved by an updated analysis of the gamma-ray data near the GC that includes the 4FGL-DR4 sources.
\item
The uncertainties in the diffuse emission model affect the detectability and the properties of pointlike sources.
In particular, many sources near the GC in the 4FGL-DR4 catalog have flags, which denote different problems either with detection of the sources or with reconstruction of their properties (the corresponding effect on the SED can be significant, as discussed in Appendix~\ref{app:msp_syst}).
\item
The separation of the GCE from the \FBs is one of the biggest uncertainties that can have a factor of 2 to 3
effect on the GCE spectrum around a few GeV \citep{2017ApJ...840...43A}.
\een
Although the study presented in this work shows that the GCE can be resolved into individual pointlike sources consistent with MSP properties, in view of the above caveats, a detailed joint analysis of the diffuse emission and pointlike sources as well as a study of the \FBs near the GC is necessary to resolve the GCE puzzle \cite{2024MNRAS.530.4395S}. 
A significant progress can be also made by multiwavelength searches for counterparts of gamma-ray sources, e.g., radio emission from MSPs~\cite{2016ApJ...827..143C}.

\begin{acknowledgments}
The author would like 
to thank Aakash Bhat, Toby Burnett, Francesca Calore, Silvia Manconi, and Troy Porter for important comments and discussions, 
to acknowledge support by the DFG grant MA 8279/3-1
and the use of the following software:
Astropy (\url{http://www.astropy.org}) \cite{2013A&A...558A..33A},
Matplotlib (\url{https://matplotlib.org/}) \cite{Hunter:2007}, 
pandas (\url{https://pandas.pydata.org/}) \cite{mckinney-proc-scipy-2010},
scikit-learn (\url{https://scikit-learn.org/stable/}) \cite{scikit-learn},
and TensorFlow (\url{https://www.tensorflow.org/}) \cite{tensorflow2015-whitepaper}.
\end{acknowledgments}

\appendix

\section{Systematic uncertainties}
\label{app:msp_syst}

\begin{figure}
\centering
\includegraphics[width=\onesize\columnwidth]{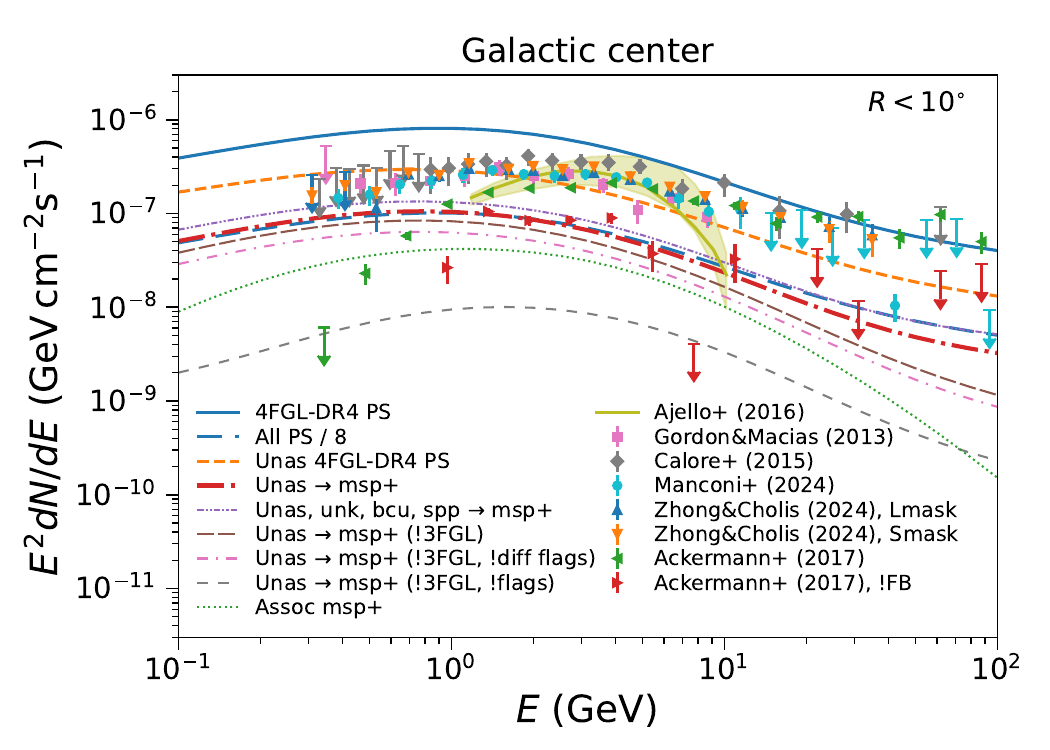}
\caption{
Effect of removing sources with analysis flags on the expected msp+ SED.
Pink sparse dash-dotted line shows the effect of removing the sources, which become insignificant as a result of changes in the diffuse emission models, which are correlated with dense gas clumps, or which have TS $< 25$ (flags 1, 6, or 13 respectively in the 4FGL-DR4 catalog \cite{2023arXiv230712546B}).
Gray sparse dashed line shows the effect of removing all flagged sources in the 4FGL-DR4 catalog.
See Fig.~\ref{fig:msp_gc_sp} for the definition of the other lines and points.
}
\label{fig:msp_flags}
\end{figure}

In this appendix we check the dependence of the predictions for the msp+ component among unassociated sources
 on the choice of the classification algorithm, uncertainties in the background diffuse emission, classification with or without bcu and spp sources, the use of weighting to account for differences in the distributions of associated and unassociated sources,
 and the effect of masking the Galactic plane within $|b| < 2^\circ$.
 
In Fig.~\ref{fig:msp_flags} we show the effect of removing sources with analysis flags.
The SED for the msp+ component of unassociated sources is shown by the red dash-dotted line.
The msp+ component of unassociated sources together with bcu, spp, and unk sources is shown by the purple dash-dot-dotted line.
The msp+ component of unassociated sources excluding the sources, which were already detected in the 3FGL catalog is shown by the brown dashed line.
The change is less than about 30\% for energies below 10 GeV (the drop is about a factor of 3 at 100 GeV), which means that the majority of the msp+ contribution around a few GeV in the 4FGL catalog is newly resolved compared to the 3FGL catalog.
Some sources in the catalog may be false detections due to uncertainties in the gamma-ray diffuse background model.
The pink sparse dash-dotted line shows the msp+ component of unassociated sources after removing the sources with the following flags \cite{2023arXiv230712546B} (referred to as diffuse flags below): (1) TS $<$ 25 with other model or analysis, (6) Interstellar gas clump (c sources), (13) TS $<$ 25.
Here we also remove sources detected in the 3FGL catalog (the label has ``!3FGL, !diff flags'').
The effect of these three flags is about 25\% or less for the whole energy range (compared to the line showing the contribution of msp+ sources without the sources detected in the 3FGL catalog).
Removing all flagged sources reduces the msp+ SED by almost a factor of 10 (gray sparse dashed line),
which shows that there are potentially problems with detection and reconstruction of spectra and / or positions of sources near the GC.
For comparison we also show the combined SED of all resolved, all unassociated, and all associated msp+ sources in the ROI by the solid blue, dashed orange, and  dotted green lines respectively.

\begin{figure}
\centering
\includegraphics[width=\onesize\columnwidth]{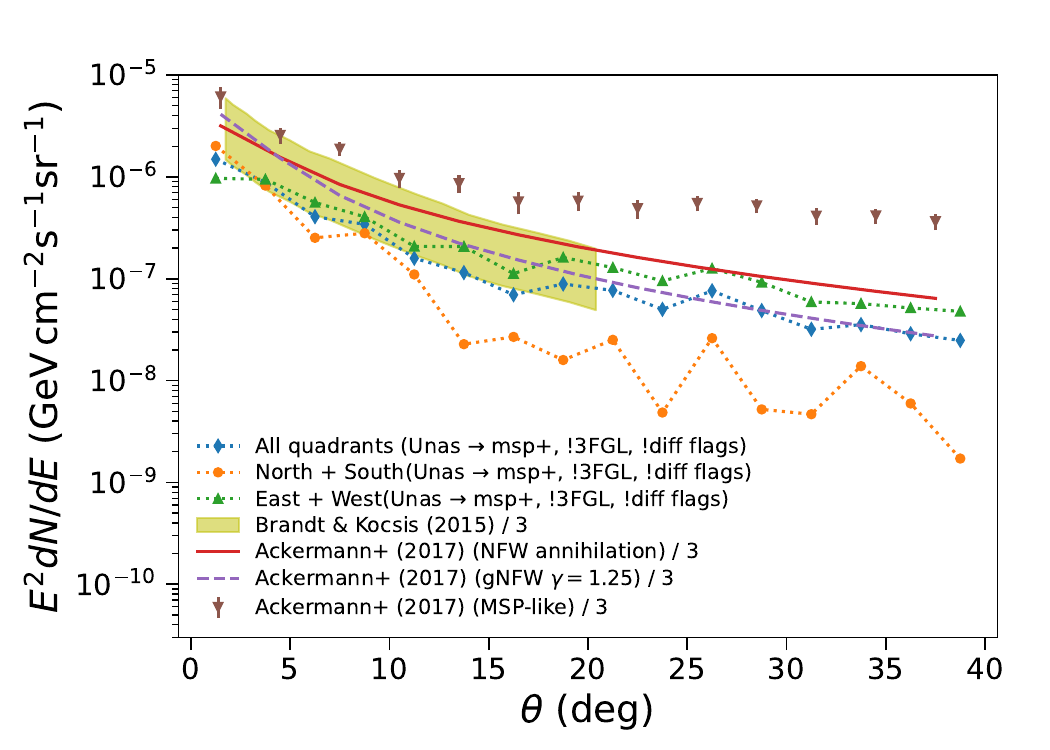}
\includegraphics[width=\onesize\columnwidth]{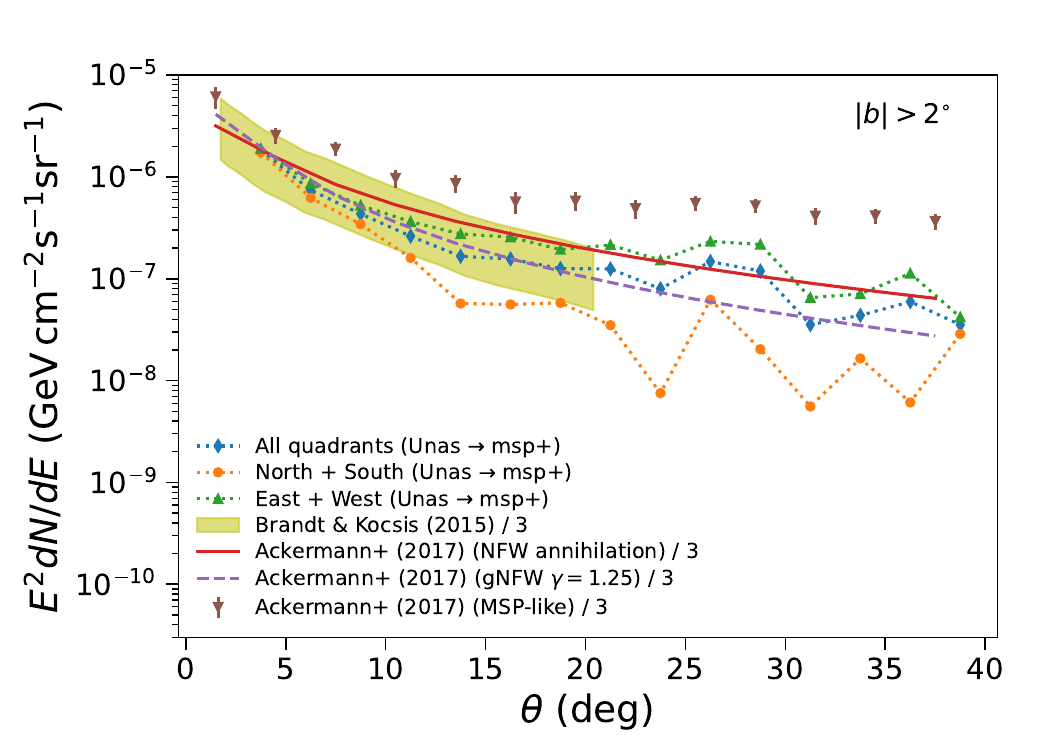}
\caption{
Top panel: effect of removing sources detected in the 3FGL catalog and sources with diffuse analysis flags 
(flags 1, 6, or 13 respectively in the 4FGL-DR4 catalog \cite{2023arXiv230712546B})
on the expected radial distribution of the msp+ component.
Lines and points are the same as in Fig.~\ref{fig:radial} but we exclude unassociated sources detected in the 3FGL catalog and
the sources with diffuse analysis flags.
Bottom panel: effect of masking sources with $|b| < 2^\circ$. 
For the definition of lines and points, see Fig.~\ref{fig:radial}.
}
\label{fig:radial_flags}
\end{figure}

The average radial profile and the profiles in north+south or east+west quadrants of the MSP-like component
after removing sources detected in the 3FGL catalog and sources with the diffuse analysis flags are shown in Fig.~\ref{fig:radial_flags}, top panel.
The profiles are similar to the profiles in Fig.~\ref{fig:radial} with the overall normalization reduced by about 40\%.
The radial profile of MSP-like sources without sources with $|b| < 2^\circ$ is shown on the bottom panel of Fig.~\ref{fig:radial_flags}.
The effect of masking the Galactic plane within $2^\circ$ is less than about 25\% up to $10^\circ$ from the GC for the east+west quadrants
(cf. Fig.~\ref{fig:radial}), 
while above $10^\circ$ from the GC the flux from the msp+ sources in the east+west quadrants is up to a factor of 2 smaller after masking the Galactic plane, which shows that at large distances from the GC the flux in the east+west quadrants is dominated by the Galactic disk sources, while within $10^\circ$ from the GC the flux is dominated by a spherical distribution of sources.

\begin{figure}
\centering
\includegraphics[width=\onesize\columnwidth]{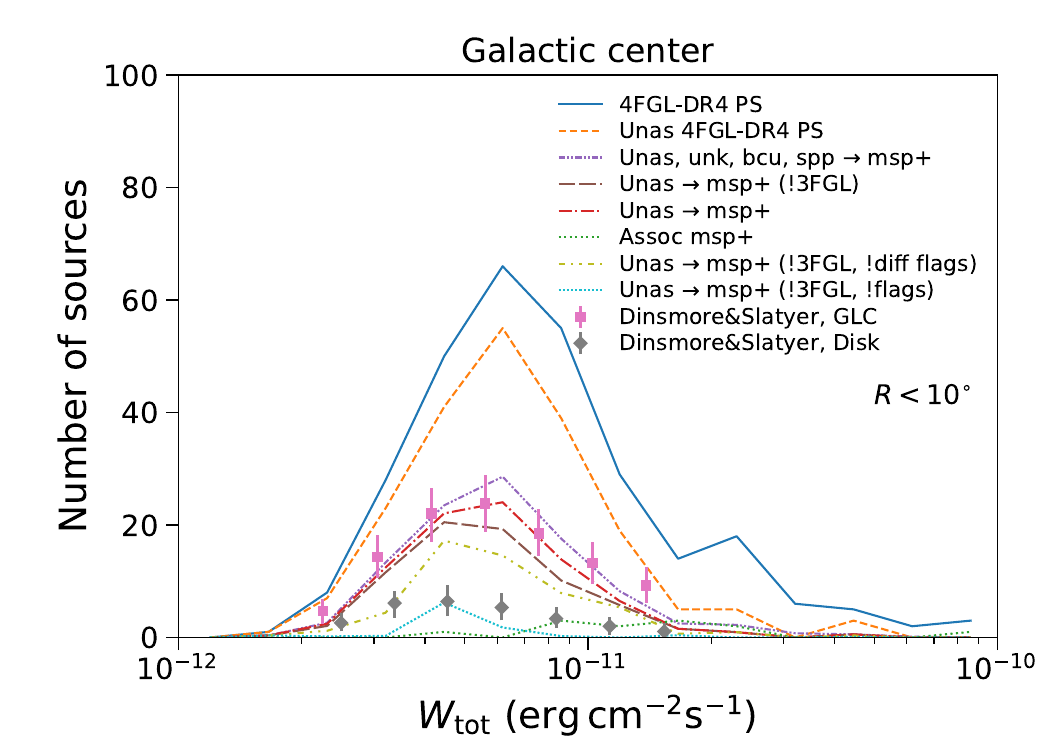}
\caption{
Effect of removing sources with diffuse analysis flags 
(sparse dash-dot-dotted yellow line)
and all analysis flags (dense dotted cyan line) on the source count distribution of the msp+ component.
The other lines and points are the same as in Fig.~\ref{fig:source_count}.
}
\label{fig:source_count_flags}
\end{figure}

The effect of removing sources with analysis flags on the source count distribution is shown in 
Fig.~\ref{fig:source_count_flags}.
Diffuse analysis flags 
(flags 1, 6, or 13 in the 4FGL-DR4 catalog \cite{2023arXiv230712546B})
mostly affect sources with energy flux below about 
$4\times 10^{-12}\ {\rm erg\, cm^{-2} s^{-1}}$, for larger fluxes the effect is less than 25\%.
Although the source count distribution after removing sources with diffuse analysis flags seems to be inconsistent with the GLC model, one should take into account the effect of removing the flagged sources on the source sensitivity model used in Ref.~\cite{2022JCAP...06..025D} to determine the predicted source count distributions of sources detected by the {\Fermi} LAT that can explain the GC excess.
Removing all sources with flags has a rather large effect on the source count distribution.
However, one should also take into account the effect on the \Fermi-LAT source detection sensitivity.
In addition, some flags are given to sources, which are unstable with respect to a change in the diffuse emission model 
(e.g., a source moves outside of the 95\% containment ellipse), but the sources themselves do not disappear.
As a result, removing sources with all analysis flags is likely a too conservative approach to estimate the contribution of pointlike sources in the ROI.

\begin{figure}
\centering
\includegraphics[width=\onesize\columnwidth]{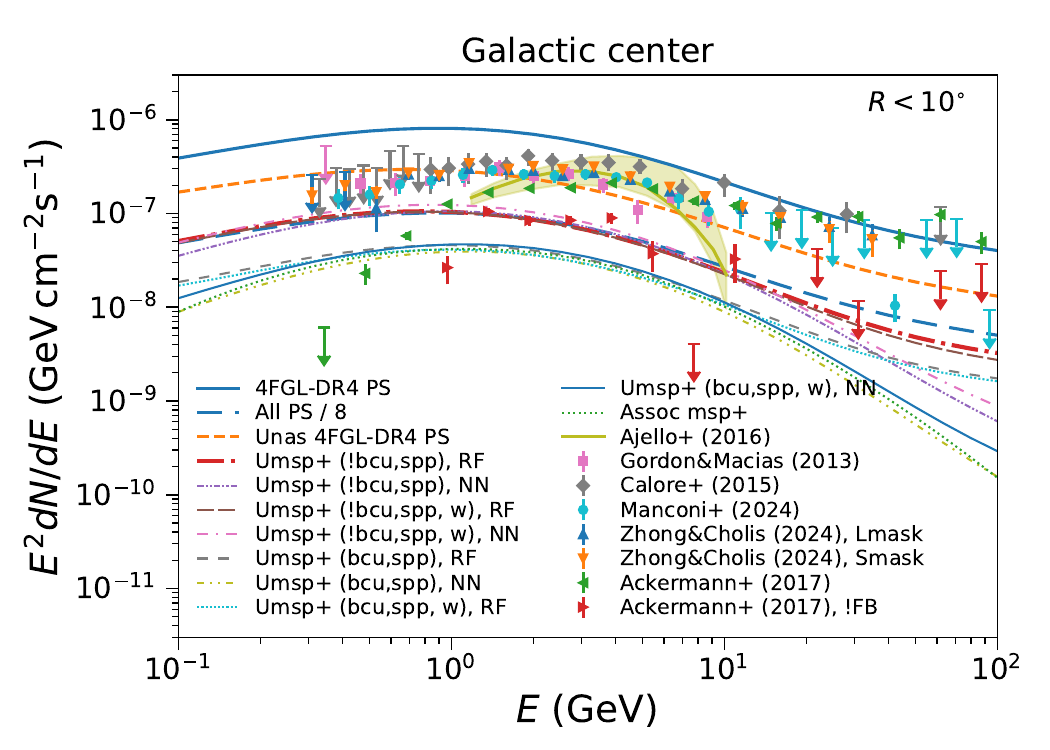}
\includegraphics[width=\onesize\columnwidth]{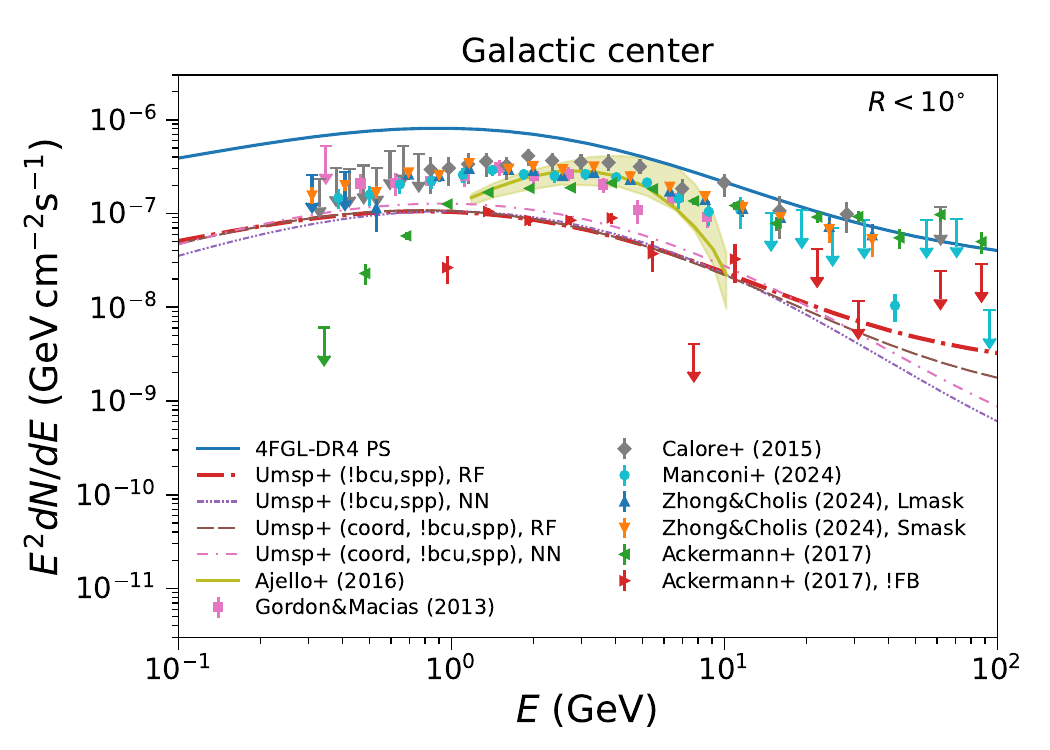}
\caption{
Top panel:
effect of the choice of the classification algorithm (RF or NN in labels), of excluding or including bcu and spp sources in training (``!bcu, spp'' or ``bcu, spp'' in labels), and of using unweighted or weighted loss to account for differences in the distribution of associated and unassociated sources (weighted loss is shown by ``w'' in labels)
on the msp+ contribution to the SED of unassociated sources.
``Umsp+'' shows the contribution of the msp+ class to the unassociated sources.
Bottom panel: effect of inclusion of coordinate features in training and predictions.
The predictions obtained with coordinate features have ``coord'' in labels.
See Fig.~\ref{fig:msp_gc_sp} for the definition of the other lines and points.}
\label{fig:msp_many_cats}
\end{figure}

In the top panel of Fig.~\ref{fig:msp_many_cats} we show the effect of several choices in classification algorithms
on the SED of the msp+ component of the unassociated sources (denoted as ``Umsp+'' in the labels).
The classification with RF and NN algorithms is denoted by ``RF'' and ``NN'' in the labels respectively.
In order to account for the differences in the distributions of associated and unassociated sources, 
we introduce weights proportional to the ratio of probability distribution functions of unassociated to associated sources
\citep{2023RASTI...2..735M}.
Training with associated sources weighted by the corresponding weights is shown by ``w'' in the labels.
The choice of the classification algorithms and the use of weighting results in changes of less than 20\% 
for the predicted SED of the msp+  component in the unassociated sources
for energies between 1 and 10 GeV.
The last two effects that we test are the inclusion of bcu and spp classes and the use of coordinate features in training and predictions.
The definition of the groups, when bcu and spp classes are included in training, as well as predictions for the overall number of unassociated and unk sources attributed to the six classes
are shown in Table~\ref{tab:pred_no_coord}.
The corresponding SEDs for the msp+ component of unassociated sources when the bcu and spp classes are included in training (``bcu,spp'' in labels) are a factor of 2 to 3 smaller than the SEDs for training without bcu and spp classes (``!bcu,spp'' in labels).
The main reason for the reduction is that when bcu and spp sources are included in training, then a significant fraction of unassociated sources is also attributed to these new classes, which reduces the expected number of MSP-like sources.
This reduction can also be seen in the expected number of msp+ sources.
The expected number of msp+ sources among unas and unk sources over the whole sky is reported in Table~\ref{tab:pred_unas_bcu_spp_unk} (for the no bcu and spp case) and in Table~\ref{tab:pred_no_coord} (when bcu and spp sources are included in training).
In the bottom panel of Fig.~\ref{fig:msp_many_cats} we show that adding coordinate features in training and classification has
little effect on the predicted SED from MSP-like sources.
The SEDs derived with coordinate features have ``coord'' in labels.

\begin{table*}
\centering
\footnotesize
\begin{tabular}{llllllll}
\hline
 & Class label & Physical classes & N assoc & RF assoc & NN assoc & RF unas,unk & NN unas,unk \\
\hline
1 & spp+ & nov, spp & 130 & $125.6\pm 12.2$ & $136.3\pm 35.1$ & $366.9\pm 21.2$ & $406.2\pm 82.8$ \\
2 & fsrq+ & fsrq, nlsy1 & 827 & $827.1\pm 29.6$ & $812.6\pm 69.7$ & $205.7\pm 10.2$ & $186.4\pm 30.5$ \\
3 & psr+ & snr, hmb, pwn, psr, gc & 219 & $208.3\pm 16.3$ & $222.0\pm 39.8$ & $157.0\pm 12.7$ & $164.0\pm 28.7$ \\
4 & msp+ & msp, lmb, glc, gal, sfr, bin & 244 & $244.3\pm 17.2$ & $243.3\pm 23.8$ & $271.3\pm 14.1$ & $265.9\pm 20.8$ \\
5 & bcu+ & sey, bcu, sbg, agn, rdg & 1698 & $1693.3\pm 33.4$ & $1718.1\pm 112.0$ & $1244.7\pm 21.7$ & $1253.1\pm 43.7$ \\
6 & bll+ & bll, ssrq, css & 1498 & $1514.8\pm 35.0$ & $1481.0\pm 119.8$ & $329.3\pm 10.1$ & $299.3\pm 42.9$ \\
\hline
\end{tabular}
\caption{Definition of classes and prediction for unassociated sources together with the unk physical class in case bcu and spp sources
are included in training \citep{2023RASTI...2..735M}.}
\label{tab:pred_no_coord}
\end{table*}

\section{Contribution of young pulsars to GCE}
\label{app:psr_contribution}

\begin{figure}
\centering
\includegraphics[width=\onesize\columnwidth]{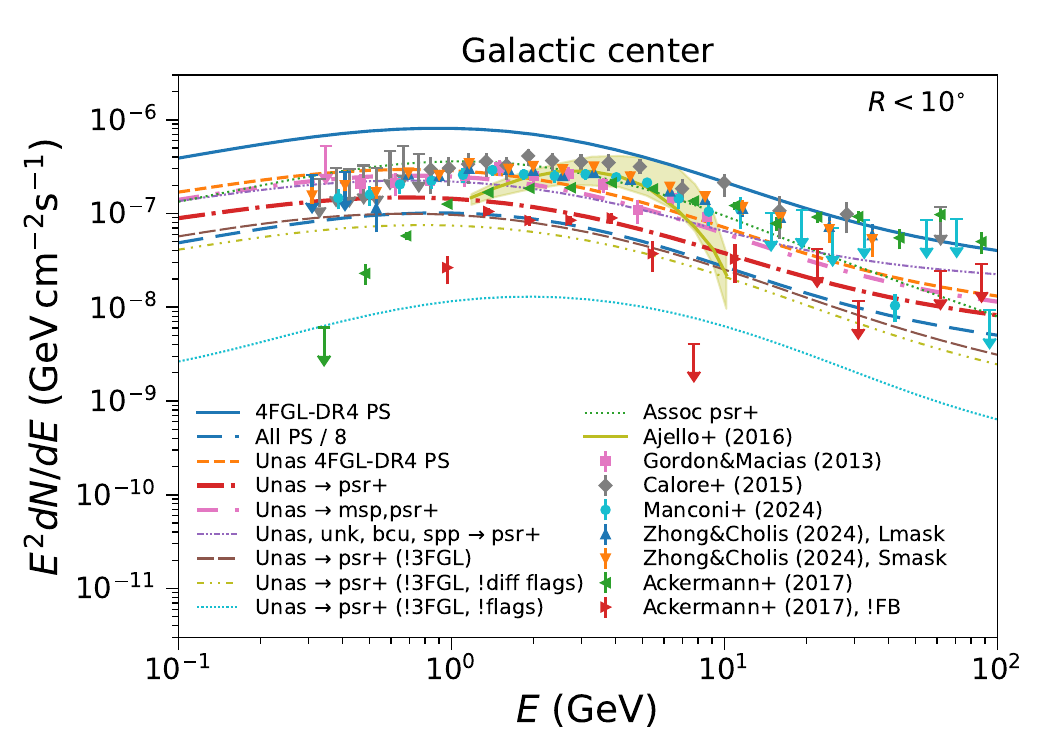}
\caption{
Contribution of the psr+ class to the SED of unassociated sources within $10^\circ$ from the GC.
Yellow sparse dash-dot-dotted line shows the combined contribution of msp+ and psr+ sources.
The other lines are either the same as in Figs.~\ref{fig:msp_gc_sp} and \ref{fig:msp_flags} or the msp+ contribution is replaced by the psr+ contribution.}
\label{fig:psr_flags}
\end{figure}

A population of young pulsars has been proposed as one of the possible explanations of the GCE 
\citep{2015arXiv150402477O}.
In this appendix we consider the contribution of the psr+ component among the unassociated sources 
to the gamma-ray flux near the GC.
Fig.~\ref{fig:psr_flags} shows the SED for the psr+ component of unassociated sources within 
$10^\circ$ from the GC (red dash-dotted line).
The contribution from the psr+ class is comparable to the msp+ component (red dash-dotted line in 
Figs.~\ref{fig:msp_gc_sp}, \ref{fig:msp_flags}, and \ref{fig:msp_many_cats}).
We also show the combined contribution from psr+ and msp+ classes (shown by the purple sparse dash-dotted line in 
Fig.~\ref{fig:psr_flags}).
The combined psr+ and msp+ SED is now above the possible GCE SED if the \FBs are taken into account
and just below the GCE, where the \FBs are not modeled or modeled with the same intensity as at high latitudes.
As in Fig.~\ref{fig:msp_flags}, we also check the effect of excluding the 3FGL sources as well as excluding sources with flags.
Excluding 3FGL sources results in the reduction of the psr+ SED by up to 40\%, which is slightly larger than the corresponding effect of
up to 25\% (Fig. \ref{fig:msp_flags}) for the msp+ class.
Removing sources with diffuse analysis flags (1, 6, and 13) gives an additional reduction of about 20\%.
Interestingly, the combined SED of associated psr+ sources in this ROI 
(green dotted line in Fig.~\ref{fig:psr_flags})
is about a factor of 2 larger than the SED of the psr+ component of unassociated sources, while the SED of associated msp+ sources (green dotted line in Fig.~\ref{fig:msp_flags}) is about a factor of 2 smaller than the msp+ component of unassociated sources.

\begin{figure}
\centering
\includegraphics[width=\onesize\columnwidth]{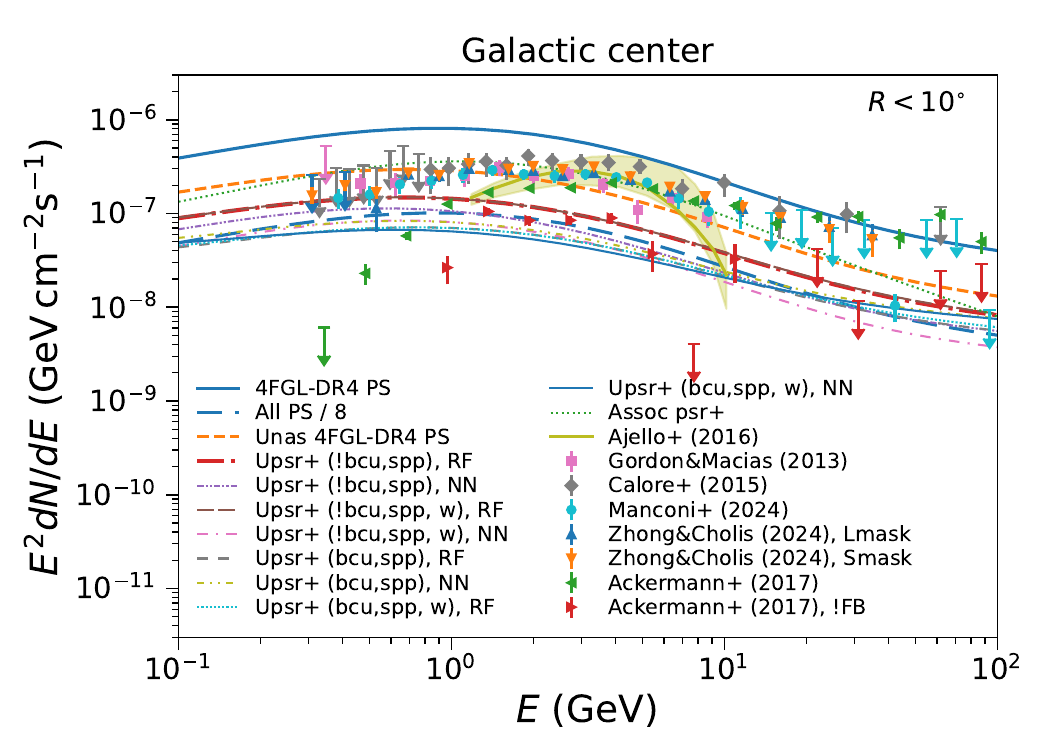}
\caption{
Effect of the choice of the classification algorithm, of excluding or including bcu and spp sources in training, and of using unweighted or weighted loss 
on the psr+ contribution to the SED of unassociated sources.
For the definition of lines see Figs.~\ref{fig:msp_gc_sp} and \ref{fig:msp_many_cats}
where the msp+ contribution is replaced by the psr+ contribution (when relevant).
}
\label{fig:psr_cats}
\end{figure}

In Fig.~\ref{fig:psr_cats} we test the effect of the choice of classification algorithm (RF vs NN), weighted or unweighted training, and including or not including bcu and spp classes in training. 
In this case, excluding bcu and spp sources from training leads up to a factor of 2 reduction of the expected
psr+ SED.

\begin{figure}
\centering
\includegraphics[width=\onesize\columnwidth]{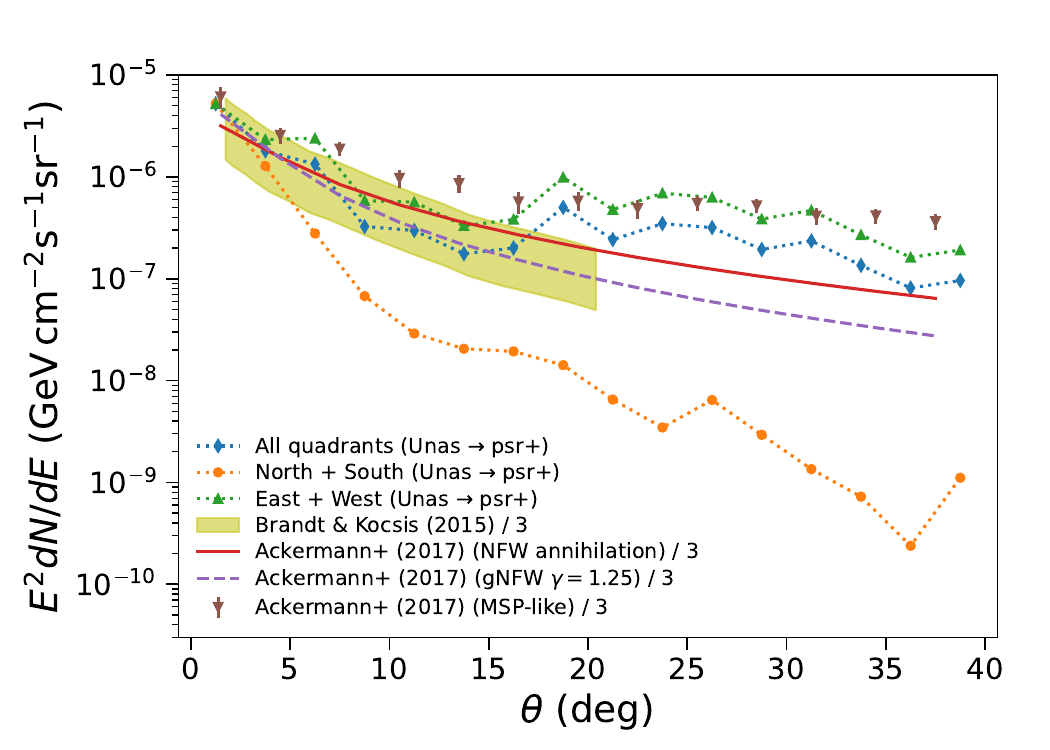}
\caption{
Contribution of the psr+ class to the radial intensity profile of unassociated sources at 2 GeV.
The lines are defined as in Fig.~\ref{fig:radial} with the msp+ contribution replaced by the psr+ contribution (when relevant).}
\label{fig:radial_psr}
\end{figure}

The radial profile of the psr+ component is shown in Fig.~\ref{fig:radial_psr}.
In this case the averaged profile is similar to the NFW DM annihilation profile. 
However, in the east+west (north+south) quadrants the psr+ profile is flatter (steeper) than the NFW annihilation profile.
This is expected for the distribution of young pulsars, since they cannot travel far from the locations where they are created
due to relatively short lifetime. As a result, the young pulsars are distributed close to the Galactic plane (see also Fig.~\ref{fig:map}).
Thus, although, the SED of the psr+ component is comparable to the msp+ component and to the expected GCE SED (after excluding a possible contribution from the \FBs), the psr+ sources have a distribution along the Galactic plane, which is different from the more spherical distribution of the gamma-ray excess around the GC as well as the distribution of the msp+ component.



\bibliography{msp_multi_gce_bibl}

\end{document}